\documentclass[twocolumn,aps,prl]{revtex4}

\usepackage{graphics}% Include figure files
\usepackage{graphicx}% Include figure files
\usepackage{dcolumn}% Align table columns on decimal point
\usepackage{bm}% bold math
\usepackage{amsmath,amssymb}

\newcommand{\de}{\partial}
\newcommand{\sech}[1]{\textrm{sech}\left(  #1\right)}
\newcommand{\eq}[2]{\begin{equation} \label{#1} #2 \end{equation}}
\newcommand{\eps}{\epsilon}

\newcommand{\sgn}{\textrm{sgn}}

\newcommand{\etal}{{\em et al.}}

\newcommand{\cc}{\textrm{c.c.}}

\newcommand{\JJ}{\mathbf{J}}
\newcommand{\pp}{\mathbf{p}}

\newcommand{\EE}{\mathbf{E}}

\newcommand{\vv}{\mathbf{v}}
\newcommand{\rr}{\mathbf{r}}

\newcommand{\FF}{\mathbf{F}}
\newcommand{\AAA}{\mathbf{A}}

\begin{document}

%\draft

% Change notation for vectors, should not be mixed with commutators

\title{Terahertz relativistic spatial solitons in doped graphene metamaterials}
\author{Haiming Dong$^{1}$, Claudio Conti$^{2}$ and Fabio Biancalana$^{1}$}
\affiliation{$^{1}$Max Planck Institute for the Science of Light, G\"{u}nther-Scharowsky-Str. 1, Bau 26, 91058 Erlangen, Germany
\\$^{2}$Dep. Mol. Med. and CNR-ISC, Univ. Sapienza, P.le A. Moro 5, 00185 Rome, Italy}
\date{\today}

\begin{abstract}
We propose an electrically tunable graphene-based metamaterial showing a large nonlinear optical response at THz frequencies,
which we calculate analytically for the first time to our knowledge and arises from the intraband current.
The structure sustains a novel type of stable two-dimensional spatial solitary wave, a relativistic version of the Townes soliton.
These results can be also applied to any material exhibiting a conical dispersion with massless Dirac fermions.
\end{abstract}

\maketitle

\paragraph{Introduction ---} Graphene is a two-dimensional, one-atom thick allotrope of carbon that has
been in the spotlight since its experimental discovery and isolation in 2004 \cite{novoselov1},
and can be considered a unifying bridge between low-energy condensed matter physics and
quantum field theory, as its two-dimensional quasi-electrons behave like massless
``relativistic'' Dirac fermions, very similarly to electrically charged 'neutrinos' \cite{geim1,castroneto}.

Graphene holds the promise for building advanced nano-electronic devices, due to
its unconventional electronic properties \cite{castroneto}. Furthermore, it also exhibits very unique
optical properties, especially in the terahertz (THz) frequency range.
To date, novel photonic devices, such as THz devices \cite{thz},
optical modulators \cite{modulator}, photodetectors \cite{detector}
and polarizers \cite{polarizer} were successfully realized.

In recent years, the huge and largely unexplored potential of graphene
for nonlinear optical applications has been outlined. An extremely strong nonlinear optical response in the THz regime
has been investigated \cite{wright,ishikawa}. Preliminary experimental results include
ultrafast saturable absorption \cite{saturable1,saturable2} and the
observation of strong four-wave mixing \cite{hendry}, which are the building blocks of nonlinear optics \cite{agrawalbook}.
Specifically, four-wave mixing implies the existence of modulational instability and optical
solitons in graphene, a significant topic that has not been previously
investigated.

In this paper, we follow the footsteps of a series of seminal papers
by Mikhailov (see Refs. \cite{mikhailov2}) based on the
semiclassical kinetic theory, and we derive {\em
analytically} the intraband optical current of a doped layer of
graphene. We prove that this theory is also
consistent with the more precise quantum approach of Ishikawa
\cite{ishikawa}, based on the Bloch equations derived from the
single-electron Dirac equation. For excitation
frequencies in the THz gap, and neglecting the interband
transitions (valid for photons below the
Fermi energy of doped layers), the two approaches give the same
result for the intraband current. We apply the
above results to describe self-focusing of two dimensional
Townes-like solitons in an electrically tunable metamaterial made of several layers of
doped graphene, interspaced by layers of silica and silicon with thickness much smaller than the wavelength of the incident light.

\begin{figure}
\includegraphics[width=8cm]{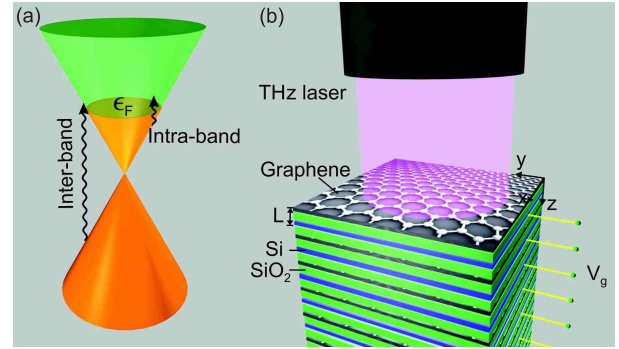}
\caption{(a) Graphene conical dispersion with doping.
Intraband and interband optical transitions are shown. (b)
Geometry of the proposed multilayer metamaterial. The structure is
made of graphene-silica-silicon layers, with total
thickness $L$, much smaller than the wavelength of the THz beam.
Each layer of graphene is doped by a gate voltage V$_{\rm g}$. \label{fig1}}
\end{figure}

\paragraph{Background ---} We consider an electrically doped
graphene system with a positive gate voltage V$_{\rm
g}$. As shown in Fig. \ref{fig1}a, the electron energy dispersion in the conduction band is
given by the Dirac spectrum, $\eps_{\pp}=v_{\rm F}p$, where
$p\equiv|\pp|=\sqrt{p_{x}^{2}+p_{y}^{2}}$ is the total momentum, $\pp\equiv(p_{x},p_{y})$, and $v_{\rm F}\simeq c/300$ if the Fermi velocity with
$c$ the vacuum light speed. The Fermi energy $\epsilon_{\rm F}$ can be
largely controlled by the voltage V$_{\rm
g}$ perpendicularly applied to the graphene-SiO$_2$-Si multilayer (Fig. \ref{fig1}b). The
velocity operator for the quasi-electrons is given by $\vv\equiv\nabla_{\pp} \eps_{\pp}=v_{\rm
F}\pp/p$.
The electron momentum distribution $f_{\pp}(\rr,t)$ in
the collisionless approximation is solution of the Boltzmann-Vlasov
kinetic equation: \eq{boltz1}{\de_{t}f_{\pp}(\rr,t)+\vv\cdot\nabla
f_{\pp}(\rr,t)+ \FF\cdot\nabla_{\pp}f_{\pp}(\rr,t)=0,} where $\rr$
and $t$ are respectively space and time coordinates, and $\FF\equiv
-e\EE$ is the force due to the electric field $\EE(\rr,t)$
($e>0$ here and in the following). Due to Jeans'
theorem \cite{jeans}, any function of the constants of motion
is a solution of the Boltzmann-Vlasov equation, and
assuming for simplicity homogeneity in the
$(x,y)$ plane (i.e. $\nabla f_{\pp}=0$), an exact solution of Eq.
(\ref{boltz1}) is the Fermi-Dirac distribution at
temperature $T$ for negligible interband transitions, namely
$f_{\pp}(t)=\mathcal{F}_{T}\left[p_{x}-p_{0,x}(t),p_{y}-p_{0,y}(t)\right]$,
where
$\mathcal{F}_{T}\left[p_{x},p_{y}\right]\equiv\left[1+\exp\left\{\left(\eps_{\pp}
-\eps_{\rm F}\right)/(k_{\rm B}T)\right\}\right]^{-1}$, $k_{B}$ is the Boltzmann constant,
$\pp_{0}(t)\equiv-e\AAA(t)$ is the electron momentum transferred
by the radiation field, $\AAA(t)=-\int\EE(t)dt$ is the
vector potential, and $\eps_{\rm F}$ corresponds to an
electron surface density $n_{\rm s}\equiv \eps_{\rm
F}^{2}/(\hbar^{2}v_{\rm F}^{2}\pi)$.

\paragraph{Calculation of total intraband current ---}  The total electric current is
given by $\JJ\equiv-\frac{g_{\rm s}g_{\rm
v}}{(2\pi\hbar)^{2}}e\int\vv f_{\pp}d\pp=-\frac{g_{\rm s}g_{\rm
v}}{(2\pi\hbar)^{2}}ev_{\rm
F}\int(\pp/p)\mathcal{F}_{T}[\pp-\pp_{0}(t)]d\pp$, where $g_{\rm
s}=2$ and $g_{\rm v}=2$ are respectively the spin and valley
degeneracy factors. This gives the intraband current that
is responsible for the strong THz nonlinearity of graphene, as
demonstrated by Mikhailov \cite{hendry,mikhailov2}.

More recently, Ishikawa \cite{ishikawa} introduced Bloch-like
equations deduced from the one-electron Dirac equation. In
his formalism, he starts from the Weyl equation for the charged
neutrino, $i\sigma^{\mu}\tilde{\de}_{\mu}\mathbf{\psi}=0$, where
$\tilde{\de}_{\mu}\equiv\left(v_{\rm
F}^{-1}\de_{t},\nabla_{\bot}\right)$ is the pseudorelativistic
derivative, $\nabla_{\bot}\equiv(\de_{x},\de_{y})$, and
$\sigma^{\mu}\equiv(\sigma^{0},\mathbf{\sigma})$ is the Pauli
matrices vector, and $\sigma^{0}$ is the $2\times 2$ identity
matrix. Expanding into space and time variables, this translates
into the wave equation for the electronic spinor with momentum $p$,
which reads $i\hbar\de_{t}\mathbf{\psi}_{p}=v_{\rm F}
(\mathbf{\sigma}\cdot\mathbf{p})\mathbf{\psi}\equiv\hat{H}_{0}\mathbf{\psi}_{p}$,
with the momentum operator $\mathbf{p}\equiv -i\hbar\nabla_{\bot}$.
The interacting theory is directly implemented via the minimal
substitution: $i\hbar\de_{t}\mathbf{\psi}_{p}=v_{\rm
F}\left(\mathbf{\sigma}\cdot\left[\mathbf{p}+
\frac{e}{c}\mathbf{A(t)}\right]\right)\mathbf{\psi}_{p}$. Without
loss of generality, we can assume that the radiation field is
polarized linearly along the $x$ direction, parallel to the graphene plane. The Weyl equation for one
electron is therefore
%\begin{widetext}
\eq{weyl1}{i\hbar\de_{t}\mathbf{\psi}_{p}=v_{\rm F}\left[\begin{array}{cc}0 &
\left(p_{x}+\frac{e}{c}A\right)-ip_{y} \\ \left(p_{x}+\frac{e}{c}A\right)+ip_{y}
& 0\end{array}\right]\mathbf{\psi}_{p}.}
%\end{widetext}
From Eq. (\ref{weyl1}), one can construct
the single-electron current $\mathbf{J}_{\rm 1,p}\equiv-ev_{\rm
F}\mathbf{\psi}_{p}^{\dagger}\mathbf{\sigma} \mathbf{\psi}_{p}$ for
a given electronic momentum $p$.
Ishikawa \cite{ishikawa} demonstrated that, for intraband transitions only ($n=-1$ and $\rho=0$ in the notation of
Ref. \cite{ishikawa}), the single-electron intraband current
reduce to \eq{intra1}
{\mathbf{J}_{1,p}=\frac{-ev_{\rm
F}}{\sqrt{(p_{x}+eA)^{2}+p_{y}^{2}}}(p_{x}+eA,p_{y}),} and
 $\JJ=\frac{g_{\rm s}g_{\rm
v}}{(2\pi\hbar)^{2}}\int\mathbf{J}_{1,p}\mathcal{F}_{T}[p_x,p_y]d\pp$.
This result agrees with the
Boltzmann-Vlasov equation approach given above.
It is important to note that, following Refs. \cite{mikhailov2}, the intraband
current dominates the interband current
for photon energies $\hbar\omega\lesssim \eps_{\rm F}$ and for
$k_{\rm B}T\ll\eps_{\rm F}$.

In this paper, we report the first analytically calculation of the total macroscopic
intraband current of a single graphene layer of thickness $d$ at low temperature ($T\rightarrow 0$). The final result for
the $x$ component (i.e. the only non-vanishing component of the
two-dimensional current) is given by
\begin{widetext}
\eq{jtot1}{J_{\rm 2D}(A)=-\frac{eg_{\rm s}g_{\rm v}v_{\rm
F}}{(2\pi\hbar)^{2}}\frac{2|p_{\rm F}+eA|}{3eA}\left\{ (p_{\rm
F}^{2}+e^{2}A^{2}) \mathcal{E}_{+}\left(\frac{4eAp_{\rm F}}{(p_{\rm
F}+eA)^{2}}\right)- (p_{\rm F}-eA)^{2}
\mathcal{E}_{-}\left(\frac{4eAp_{\rm F}}{(p_{\rm F}+eA)^{2}}\right)
\right\},}
\end{widetext}
where we have defined the elliptic integrals
$\mathcal{E}_{\pm}(x)\equiv\mathcal {E}_{\pm}(\frac{\pi}{2}|x)$, with
$\mathcal{E}_{\pm}(\theta|x)\equiv\int_{0}^{\theta}
(1-x\sin^{2}\theta)^{\pm 1/2}d\theta$.

\begin{figure}
\includegraphics[width=8cm]{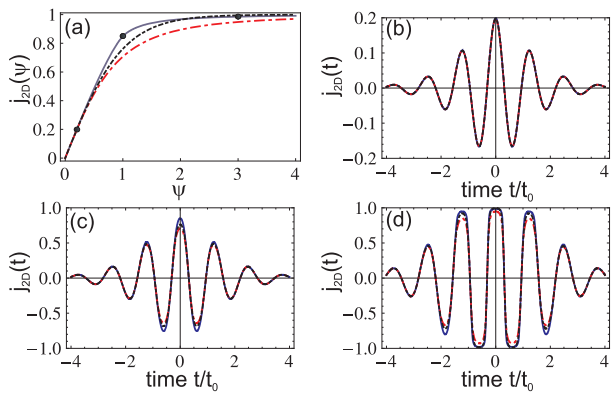}
\caption{(a) Plot of the current $j_{\rm 2D}(\psi)$
(blue solid line), its hyperbolic tangent approximation (black dashed-dotted line) and its relativistic approximation (red dashed line).
(b,c,d) Plots of analytical intraband current $j_{\rm 2D}(t)$ when $\psi(t)=\psi_{0}\sech{t/t_{0}}\cos(5t/t_{0})$ (blue solid line),
for $\psi_{0}=0.2$, $1$ and $3$ respectively, corresponding to the three black dots in (a). The lines for the tanh (black dashed-dotted) and the relativistic (red dashed)
approximations are also shown.
\label{fig2}}
\end{figure}

By defining the dimensionless variable $\psi\equiv eA/p_{\rm F}$, where $p_{\rm F}\equiv\eps_{\rm F}/v_{\rm F}$ is the Fermi momentum, and electric monochromatic fields are scaled
with the reference field $E_{0}\equiv\omega\eps_{\rm F}/(v_{\rm F}e)$.
Fig. \ref{fig2}(a) shows the dimensionless quantity $j_{\rm
2D}(\psi)\equiv J_{\rm 2D}/j_{\rm F}$ (blue solid line) [where $j_{\rm F}\equiv-ev_{\rm
F}p_{\rm F}^{2}/(\pi\hbar^{2})$ is the elementary Fermi current] which in terms of the $\psi$ variable is given by
\begin{widetext}
\eq{jtot2}{j_{\rm 2D}(\psi)=\frac{2}{3\psi}|1+\psi|\left\{(1+\psi^2)
\mathcal{E}_{+}\left(\frac{4\psi}{(1+\psi)^{2}}\right)-
(\psi-1)^{2}\mathcal{E}_{-}\left(\frac{4\psi}{(1+\psi)^{2}}\right)\right\},}
\end{widetext} showing the strong
intrinsic nonlinear behavior of graphene layers when excited
by THz radiation. Interestingly, $j_{\rm 2D}(\psi)$ can be roughly
approximated in several ways, depending on the specific application.
For instance, a hyperbolic tangent function approximation, $j_{\rm
2D}(\psi)\simeq\tanh(\psi)$, is useful for several estimates,
for example in the calculation of the linear intraband conductivity, giving the correct
$\sigma_{\rm intra}=e^2\eps_{\rm F}/(\pi\hbar^2\omega)$ with
$\omega$ being the radiation frequency. We note that the latter
function has exactly the same asymptotic behavior of $j_{\rm 2D}$
for $\psi\rightarrow 0$ and $\psi\rightarrow\infty$. However the approximation might fail for more precise
estimates around the most nonlinear region, namely $\psi=1$.
Here, however, we prefer to use the less precise but more tractable approximation
$j_{\rm 2D}(\psi)\simeq\psi/\sqrt{1+\psi^{2}}$, which shows also the pseudo-relativistic
nature of the optical nonlinearity treated here. This expression will allow us to treat the
transition from real fields to envelopes in a straightforward way in the framework of the paraxial approximation, since the
Taylor expansion of Eq. (\ref{jtot2}) does not work well due to its fictitious singularity at $\psi=0$.

In Fig. \ref{fig2}(a) $j_{\rm 2D}(\psi)$ and its two approximated versions are shown for comparison.
In Figs. \ref{fig2}(b,c,d) we show the analytically calculated current $j_{\rm 2D}(t)$ for an
example of pulsed excitation $\psi(t)=\psi_{0}\sech{t/t_{0}}\cos(5t/t_{0})$, where $t_{0}$ is the pulse width, for
different values of the light field amplitude $\psi_{0}$, showing
the strong nonlinear temporal dependence of the intraband current. Curves obtained by using the tanh and the relativistic approximations above
are also shown for comparison.

A quick estimate of the third-order susceptibility $\chi^{(3)}_{\rm gr}$ is given by expanding $j_{\rm 2D}$ in powers of $\psi$ up to the third order: $j_{\rm 2D}\simeq \psi-\psi^{3}/8$, giving the nonlinear third order intraband current $J_{\rm 2D}^{(3)}(E)=[eE/(2\omega p_{\rm F})]^{3}[ev_{\rm F}p_{\rm F}^{2}/(\pi\hbar^{2})]=\omega\eps_{0}\chi^{(3)}_{\rm gr}E^{3}$. The order of magnitude of such susceptibility for a monochromatic wave is thus given by $\chi^{(3)}_{\rm gr}=e^4v_{\rm F}^2/(8\pi \eps_0\hbar^2\omega^4\eps_{\rm F} d) \sim 10^{8}\div 10^{14}\chi^{(3)}_{\rm silica}$, depending on the specific parameters used. Note that $\chi^{(3)}_{\rm gr}\sim\omega^{-4}$, so the intraband nonlinearity rapidly decreases when increasing the frequency.
This is consistent with the estimates given by Mikhailov \cite{hendry,mikhailov2}. Such estimates, although relevant to retrieve the orders of magnitude involved, do not capture the full complexity of the nonlinearity, which is not a simple Kerr nonlinearity. However, such a large third order coefficient
places the graphene nonlinearity in the same category of the resonant nonlinear effects, such as two-level systems \cite{boydbook} or the excitonic nonlinearity \cite{biancalanasmyrnov}, but with the great advantage that the bandstructure of graphene is {\em always resonant} to optical excitations.

\paragraph{Wave equation for a single graphene layer ---} The equation that regulates light propagation in presence
of a single graphene layer is the conventional
macroscopic wave equation $c^{2}\eps_{0}\Box A=J_{\rm 3D}(A)$, where
$\Box\equiv (\eps_{\rm s}/c^{2}) \de_{t}^{2}-\nabla^{2}$, with
$\eps_{\rm s}$ is the substrate dielectric function at the selected
frequency $\omega$, and $A(\rr,t)$ is the 3D vector potential. The
current circulates in a very thin layer of thickness $d\simeq 0.34$ nm \cite{thickgraphene}. Thus we can model the 3D
current with a rectangular function, with the single layer centered
at $z=0$: $J_{\rm 3D}(\rr,t)=J_{\rm 2D}(x,y,t)R(z)/d$,
where $R(z)\equiv\left\{\sgn(z+d/2)+\sgn(d/2-z)\right\}/2$, where $\sgn(x)$ is the sign function,
normalized like $\int_{-\infty}^{+\infty}R(z)dz=d$.

\paragraph{Average medium theory of graphene metamaterial ---}
In order to observe THz spatial solitons, we consider a doped
graphene metamaterial as shown in Fig. 1(b). The system that we propose is
a periodic multilayer based on graphene-silica-silicon layers, with total thickness $L$ [see Fig. \ref{fig1} (b)] where SiO$_2$ and Si are both transparent for THz and
optical frequencies \cite{transparencies}. Each graphene layer is doped with an electronic density $n_{\rm s}$ by applying a gate
voltage V$_{\rm g}$. The size $L$ of the elementary cell is assumed to be much smaller than the incident monochromatic wavelength, $L\ll\lambda$. This means that we can use an average medium approach by expanding the dimensionless vector potential in its Fourier components $\phi_{m}$:
\eq{periodicity1}{\psi(x,y,z,t)=\frac{1}{2}\left[\sum_{m\in\mathbb{Z}}\phi_{m}(x,y,z)
e^{2\pi imz/L+ik_{0}z-i\omega t}+\cc\right],} where $k_{0}$ is the
linear wavenumber, and $z$ is the direction perpendicular to the layers [see Fig. \ref{fig1}(b)]. By retaining only the fundamental order in the Fourier expansion (with $\phi\equiv\phi_{0}$),
and after using the paraxial approximation for reducing the order of the $z$ derivative, one obtains:
\begin{widetext}
\eq{eqx1}{\left\{\left(\frac{\eps_{\rm
s}\omega^{2}}{c^{2}}-k_{0}^{2}\right)
+\left(\de_{x}^{2}+\de_{y}^{2}\right)+2ik_{0}\de_{z}\right\}\phi
+\left[\frac{-e^{2}\eps_{\rm F}}{\pi \eps_{0}\hbar^{2}c^{2}d}\right]j_{\rm 2D}(\phi)c_{0}=0,}
\end{widetext}
where
$\sum_{n\in\mathbb{Z}}R(z+nL)=\sum_{n\in\mathbb{Z}}c_{n}e^{2\pi
inz/L}$, $c_{n}\equiv\frac{1}{L}\int_{-L/2}^{+L/2}R(z)e^{-2\pi i
nz/L}dz$, $c_{0}\equiv\frac{1}{L}\int_{-L/2}^{+L/2}R(z)dz=d/L$.
We now use the relativistic approximation of the full current, namely $j_{\rm2D}(\phi)\simeq\phi/[\sqrt{1+|\phi|^{2}/2}]$,
inside Eq. (\ref{eqx1}).

\paragraph{Paraxial model and soliton solutions ---} After
introducing the scalings $(x,y)=x_{0}(X,Y)$, $z=z_{0}Z$,
$x_{0}\equiv [\pi\hbar^{2}c^{2}\eps_{0}\eps_{\rm s}L/(e^{2}\eps_{\rm
F})]^{1/2}$, $ z_{0}\equiv 2k_{0}x_{0}^{2}$,
$k_{0}\equiv\sqrt{\eps_{\rm s}}\omega/c$ and
$\eta\equiv\phi/\sqrt{2}$, one obtains the paraxial equation:
\eq{paraxial1}{i\de_{Z}\eta+(\de_{X}^{2}+\de_{Y}^{2})\eta-\frac{\eta}{\sqrt{1+|\eta|^{2}}}=0}
For instance, for typical parameters $\omega/(2\pi)=20$ THz,
$\eps_{\rm s}\simeq 4.5$ \cite{dielectric}, $L\simeq 2$ $\mu$m and
$n_{\rm s}\simeq 5\times10^{12}$ cm$^{-2}$, one has $\eps_{\rm
F}\simeq 259$ meV, $x_{0}\simeq15$ $\mu$m and $z_{0}\simeq 415$
$\mu$m. The scaling for the electric field is $E_{0}\equiv\omega\eps_{\rm F}/(v_{\rm F}e)\simeq 30$ MW/cm$^{2}$. For the above parameters the room temperature is a good
approximation of the above results obtained for $T=0$, since $k_{\rm B}T\simeq 25$ meV $\ll\eps_{F}$.

\begin{figure}
\includegraphics[width=9cm]{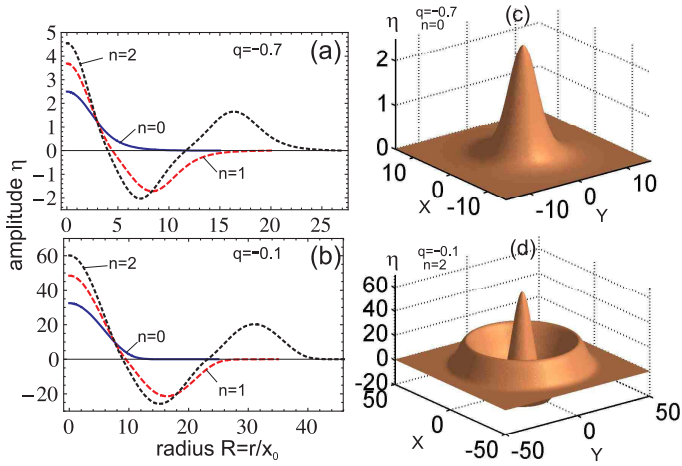}
\caption{(a) Soliton profiles of the fundamental mode ($n=0$) and two higher-order modes ($n=1$ and $n=2$) as a function of dimensionless radius $R$, for $q=-0.7$. (b) Same as (a), but for $q=-0.1$. (c,d) 3D plots of the fundamental and the 2nd-order soliton of (a,b) respectively. \label{fig3}}
\end{figure}

Passing to cylindrical coordinates one must solve the following ODE: \eq{paraxial2}{\frac{d^{2}\eta(R)}{dR^{2}}+\frac{1}{R}
\frac{d\eta(R)}{dR}-\left[q+\frac{1}{\sqrt{1+\eta(R)^{2}}}\right]\eta(R)=0,} where $q$ is a nonlinear wavenumber, and $R\equiv\sqrt{X^{2}+Y^{2}}$ is the dimensionless radius, in units of $x_{0}$.
Solutions of Eq. (\ref{paraxial2}) are Townes-like solitons (see Refs. \cite{townes,haus,sotocrespo,edmunson}) with a rather unconventional relativistic nonlinearity, which are stable in the sense of the Vakhitov-Kolokolov criterion due to the saturable type of the nonlinearity \cite{kivsharbook}. Some fundamental and higher-order soliton profiles are shown in Fig. \ref{fig3} for different values of $q$.

\paragraph{Conclusions ---}
We proposed an electrically tunable metamaterial based on graphene-silicon-silica multilayers.
We calculated the intraband current of doped graphene analytically, which dominates the electron dynamics for THz excitations.
Finally, stable Townes-like spatial solitary waves have been found to propagate in the longitudinal direction for realistic parameters.
These results pave the way of a more extensive analysis of the mathematical structure and the physical content of the graphene Bloch equations for  useful nonlinear optical applications.
Our theoretical approach is not restricted to graphene, but can be applied to all materials exhibiting a conical dispersion supporting massless Dirac fermions (for instance HgTe \cite{hgte}).

FB is funded by the German Max Planck Society for the Advancement of Science (MPG).
HD is supported by the Chinese Academy of Science (CAS).
CC acknowledges support from ERC Grant no. 201766 and the Humboldt Foundation (Germany).

\end{document}